\documentstyle[preprint,aps,epsfig,graphicx]{revtex}

\newcommand{\lsim}{\raisebox{0.3mm}{\em $\, <$} 
\hspace{-3.3mm} \raisebox{-1.8mm}{\em $\sim \,$}}

\begin{document}
\baselineskip 7.6mm
\begin{flushright}
\vglue -1.0cm
hep-ph/0202003 \\ 
\end{flushright}

\draft
\begin{center}
\Large\bf
Constraints on Neutrino Mixing Parameters By 
Observation of Neutrinoless Double Beta Decay
\end{center}
\vskip 0.5cm
\begin{center}
Hisakazu Minakata 
\footnote[1]{E-mail: minakata@phys.metro-u.ac.jp}
and Hiroaki Sugiyama
\footnote[2]{E-mail: hiroaki@phys.metro-u.ac.jp}\\
\vskip 0.2cm
{\it Department of Physics, Tokyo Metropolitan University \\
1-1 Minami-Osawa, Hachioji, Tokyo 192-0397, Japan}

\end{center}
\date{October 2001}

\vspace{-1.0cm}

\begin{abstract}
Assuming positive observation of neutrinoless double bata decay 
together with the CHOOZ reactor bound, 
we derive constraints imposed on neutrino mixing parameters, 
the solar mixing angle $\theta_{12}$ and the observable mass 
parameter $\langle m \rangle_{\beta}$ in 
single beta decay experiments. 
We show that 
0.05 eV $\leq \langle m \rangle_{\beta} \leq$ 2 eV 
at the best fit parameters of the LMA MSW solar neutrino 
solution by requiring the range of the parameter 
$\langle m \rangle_{\beta \beta}$ deduced from recently 
announced double beta decay events at 95 \% CL 
with $\pm$50 \% uncertainty of nuclear matrix elements.

\end{abstract}

\vskip 1cm

\pacs{14.60.Pq, 26.65.+t, 23.40.-s}

\newpage
\section{Introduction}

While the good amount of evidences for the neutrino mass and 
the lepton flavor mixing have been accumulated 
\cite {SKatm,solar,K2K}, 
we still lack observational indications of how large is the 
absolute mass of the neutrinos. To our understanding to date 
it may show up in only a few places, the single 
\cite {beta} or the double beta decay \cite {betabeta} experiments 
as well as future cosmological observations \cite {cosmological}. 
Other potential possibilities for hints of absolute mass of 
neutrinos include Z-burst interpretation of highest energy 
cosmic rays \cite {weiler}.

Among these various experimental possibilities the 
neutrinoless double beta decay experiments seems to have 
relatively higher sensitivities. The most stringent bound 
on effective mass parameter $\langle m \rangle_{\beta \beta}$
(see eq.~(\ref{beta1}) for definition) is now 
$\langle m \rangle_{\beta \beta} < 0.35$ eV, 
which comes from Heidelberg-Moscow group \cite{HeidelMos01}. 
Furthermore, a wide variety of proposals for future facilities 
as well as ongoing attempt with greater sensitivies 
are actively discussed. 
They include NEMO \cite{NEMO}, GENIUS \cite{GENIUS}, 
CUORE \cite {CUORE}, MOON \cite {MOON}, XMASS \cite {XMASS}, 
and EXO \cite {EXO} projects. 
These high-sensitivety experiments open the 
enlighting possibility of discovering neutrinoless double 
beta decay events, not just placing an upper bound on 
$\langle m \rangle_{\beta \beta}$ by its nonobservation.
Therefore, it is of great importance to completely understand 
what kind of informations can be extracted if such discovery 
is made. 

We discuss in this paper in a generic three flavor 
mixing framework the constraints on neutrino masses and mixing 
by positive observation (as well as nonobservation) of 
neutrinoless double beta decay.
The constraints imposed on neutrino mixing parameters by 
neutrinoless double beta decay have been discussed by 
many authors. They include the ones in early epoch \cite{early}, 
those in "modern era" in which real constraints on solar mixing 
parameters are started to be extracted \cite {MY97,dblbeta1}, and 
the ones in "post-modern era" where the analyses are performed 
in a complehensive manner in the framework of generic three flavor 
neutrino mixing \cite {dblbeta2}.

In a previous paper, we have made a final step in the series of 
analyses by proposing a way of expressing the constraints solely 
in terms of observables in single and double beta decay \cite{MS01}. 
By using the framework, we discussed the possibility of placing 
lower bound on $|U_{e3}|^2$ assuming positive observation 
in direct mass measurement in single beta decay and an upper 
limit on $\langle m \rangle_{\beta \beta}$ in double beta decay 
experiments. 
It is a natural and logical step for us to examine next 
the alternative case of positive observation of neutrinoless 
double beta decay events.

Timely enough, an evidence of the neutrinoless double beta decay 
has just been reported by Klapdor-Kleingrothaus and 
collaborators \cite{evidence}. 
Since the confidence levels of the claimed evidence are about 2 and 
3 $\sigma$ in Bayesian and Partcle Data Group methods, respectively, 
we must wait for confirmation by further data taking, or by 
other groups to conclude that neutrinos are Majorana particles. 
Nevertheless, we feel that the peak in the relevant kinematic 
region in their experiments is too prominent to be simply ignored.

As will become clear as we proceed it is essential to combine 
the constraint on $|U_{e3}|^2 = s^2_{13}$ imposed by the 
reactor experiments \cite{CHOOZ}. One of the key points in our 
subsequent discussion is that the double beta and the reactor 
bounds cooperate to produce a stringent constraint on 
absolute mass scale of neutrinos and the mixing angle 
$\theta_{12}$ which is responsible for the solar neutrino problem.

\section{Constraints from neutrinoless double beta decay}

Let us start by defining our notations. 
We use throughout this paper the standard notation of the MNS matrix
\cite{MNS}:
\begin{equation}
U=\left[
\begin{array}{ccc}
c_{12}c_{13} & s_{12}c_{13} &   s_{13}e^{-i\delta}\nonumber\\
-s_{12}c_{23}-c_{12}s_{23}s_{13}e^{i\delta} &
c_{12}c_{23}-s_{12}s_{23}s_{13}e^{i\delta} & s_{23}c_{13}\nonumber\\
s_{12}s_{23}-c_{12}c_{23}s_{13}e^{i\delta} &
-c_{12}s_{23}-s_{12}c_{23}s_{13}e^{i\delta} & c_{23}c_{13}\nonumber\\
\end{array}
\right].
\label{MNSmatrix}
\end{equation}
Using the notation, the observable in neutrinoless double beta decay 
experiments can be expressed as 
\begin{eqnarray}
\langle m \rangle_{\beta \beta}
&=& 
\left\vert 
\hskip 0.2cm
\sum^{3}_{i=1}
m_i U^2_{ei} 
\hskip 0.2cm
\right\vert \nonumber\\
&=&
\left\vert 
\hskip 0.2cm
m_1 c_{12}^2c_{13}^2 e^{-i \beta}
+ m_2 s_{12}^2c_{13}^2 e^{+i \beta}
+ m_3 s_{13}^2 e^{i(3 \gamma - 2\delta)}
\hskip 0.2cm
\right\vert,
\label{beta1}
\end{eqnarray}
where $m_i$ (i=1, 2, 3) denote neutrino mass eigenvalues, $U_{ei}$ 
are the elements in the first low of the MNS matrix, and 
$\beta$ and $\gamma$ are the extra CP-violating phases 
characteristic to Majorana neutrinos \cite {SV}, for which 
we use the convention of Ref. \cite {MY97}.

We define the neutrino mass-squared difference as
$\Delta m^2_{ij} \equiv m^2_{j} - m^2_{i}$ in this paper.
In the following analysis, we must distinguish the two 
different neutrino mass patterns, the normal ($\Delta m^2_{23}>0$) 
vs. inverted ($\Delta m^2_{23}<0$) mass hierarchies. 
We use the convention that $m_3$ is the largest (smallest) mass 
in the normal (inverted) mass hierarchy so that the angles 
$\theta_{12}$ and $\theta_{23}$ are always responsible for the 
solar and the atmospheric neutrino oscillations, respectively. 
We therefore often use the notations 
$|\Delta m^2_{23}| \equiv \Delta m^2_{atm}$ and 
$\Delta m^2_{12} \equiv \Delta m^2_{\odot}$ to emphasize 
that they are experimentally measurable quantities. 
Because of the hierarchy of mass scales, 
$\Delta m^2_{\odot}/\Delta m^2_{atm} \ll 1$, $\Delta m^2_{12}$
can be made always positive as far as $\theta_{12}$ is taken in
its full range [0, $\pi/2$]\cite {FLMP}.

In order to derive constraint on mixing parameters we need the 
classification.

\begin{equation}
\mbox{Case I:} 
\hskip 1cm 
\left\vert m_1 c_{12}^2c_{13}^2 e^{-i \beta}
+ m_2 s_{12}^2c_{13}^2 e^{+i \beta}
\right\vert
\geq m_3 s^2_{13}
\end{equation}
\begin{equation}
\mbox{Case II:} 
\hskip 1cm 
\left\vert m_1 c_{12}^2c_{13}^2 e^{-i \beta}
+ m_2 s_{12}^2c_{13}^2 e^{+i \beta}
\right\vert
\leq m_3 s^2_{13}
\end{equation}
However, examination of the case II reveals that it does not lead 
to useful bounds. Therefore, we only discuss the case I in the 
rest of this paper.

\subsection{Joint constraint by upper bounds on 
$\langle m \rangle_{\beta \beta}$ and reactor experiments}

Since we try to utilize the experimental upper bound on 
$\langle m \rangle_{\beta \beta}$, 
$\langle m \rangle_{\beta \beta} \leq 
\langle m \rangle_{\beta \beta}^{max}$, 
we derive the lower bound on $\langle m \rangle_{\beta \beta}$.
It can be obtained in the following way;

\begin{eqnarray}
\langle m \rangle_{\beta \beta} 
&\geq& c_{13}^2
\left\vert (m_1 c_{12}^2 + m_2 s_{12}^2) \cos{\beta}
- i (m_1 c_{12}^2 - m_2 s_{12}^2) \sin{\beta}
\right\vert
- m_3 s_{13}^2 \nonumber\\
&=&
c_{13}^2 \sqrt{m_1^2 c_{12}^4 + m_2^2 s_{12}^4 
+ 2 m_1 m_2 c_{12}^2 s_{12}^2 \cos{2 \beta}}
- m_3 s_{13}^2. 
\label{beta3}
\end{eqnarray}
Noticing that the right-hand-side (RHS) of (\ref{beta3}) 
has a minimum at $\cos{2 \beta}=-1$, we obtain the inequality
\begin{equation}
\langle m \rangle_{\beta \beta} \geq
c_{13}^2 
\left\vert m_1 c_{12}^2 - m_2 s_{12}^2 \right\vert
- m_3 s^2_{13}.
\label{beta4}
\end{equation}

We note that the RHS of (\ref{beta4}) is a decreasing function of 
$s^2_{13}$, and hence takes a minimum value for the maximum 
value of $s^2_{13}$ which is allowed by the limit 
placed by the reactor experiments \cite {CHOOZ}. 
We denote the maximum value as $s^2_{13}(\mbox{CH})$ throughout 
this paper. Numerically, $s^2_{13}(\mbox{CH}) \simeq 0.03$.
(While the precise value of the CHOOZ constraint actually depends 
upon the value of $\Delta m^2_{atm}$ \cite {CHOOZ}, we do not 
elaborate this point in this paper.) Using the constraint we obtain 
\begin{equation}
\langle m \rangle_{\beta \beta}^{max} \geq
\langle m \rangle_{\beta \beta} \geq
\left\vert m_1 c_{12}^2 - m_2 s_{12}^2 \right\vert - 
\left(m_3 + \left\vert m_1 c_{12}^2 - m_2 s_{12}^2 \right\vert \right)
s^2_{13}(\mbox{CH}).
\label{beta5}
\end{equation}
It can be rewritten as the bound on 
$\cos{2 \theta_{12}} = \cos{2 \theta_{\odot}}$ as 
\begin{equation}
\frac {m_2 - m_1}{m_2 + m_1} - 
\frac {\langle m \rangle_{\beta \beta}^{max} + 
m_3 s^2_{13}(\mbox{CH})}
{\frac{1}{2}(m_2 + m_1) c^2_{13}(\mbox{CH})}
\leq
\cos{2 \theta_{12}} 
\leq 
\frac {m_2 - m_1}{m_2 + m_1} + 
\frac {\langle m \rangle_{\beta \beta}^{max} + 
m_3 s^2_{13}(\mbox{CH})}
{\frac{1}{2}(m_2 + m_1) c^2_{13}(\mbox{CH})}, 
\label{bound1}
\end{equation}
where $c^2_{13}(\mbox{CH}) \equiv  1 - s^2_{13}(\mbox{CH})$.

\subsection{Joint constraint by lower bounds on 
$\langle m \rangle_{\beta \beta}$ and reactor experiments}

A positive observation of neutrinoless double beta decay 
will lead to the experimental lower bound on 
$\langle m \rangle_{\beta \beta}$, 
$\langle m \rangle_{\beta \beta} \geq 
\langle m \rangle_{\beta \beta}^{min}$, 
which we use to place new bound on neutrino 
mixing parameters. Toward the goal we note, similarly as  
(\ref{beta3}), that 
\begin{equation}
\langle m \rangle_{\beta \beta} 
\leq 
c_{13}^2 
\sqrt{m_1^2 c_{12}^4 + m_2^2 s_{12}^4 
+ 2 m_1 m_2 c_{12}^2 s_{12}^2 \cos{2 \beta}}
+ m_3 s_{13}^2,  
\label{beta6}
\end{equation}
whose RHS is maximized by taking $\cos{2 \beta} = +1$ 
and $s^2_{13} = s^2_{13}(\mbox{CH})$ in the last term 
and $c^2_{13} = 1$ in front of the square root.
(A more refined treatment entails the same excluded region.) 
One can then derive an inequality similar to (\ref{beta5});
\begin{equation}
\langle m \rangle_{\beta \beta}^{min} \leq
\langle m \rangle_{\beta \beta} \leq
\left(m_1 c_{12}^2 + m_2 s_{12}^2 \right) +  
m_3 s^2_{13}(\mbox{CH}).
\label{beta7}
\end{equation}
By rewriting (\ref{beta7})
we obtain the other upper bound on $\cos{2 \theta_{12}}$; 
\begin{equation}
\cos{2 \theta_{12}} \leq 
\frac {m_2 + m_1}{m_2 - m_1} - 
\frac {\langle m \rangle_{\beta \beta}^{min} -  
m_3 s^2_{13}(\mbox{CH})}
{\frac{1}{2}(m_2 - m_1)}. 
\label{bound2}
\end{equation}

To summarize, we have derived in this section the two kinds of 
upper bound on $\cos{2 \theta_{12}}$ 
(lower bound for $\cos{2 \theta_{12}} < 0$) 
by using the assumed experimental constraint 
$\langle m \rangle_{\beta \beta}^{min} \leq
\langle m \rangle_{\beta \beta} \leq 
\langle m \rangle_{\beta \beta}^{max}$ 
imposed by neutrinoless double beta decay experiments.

\section{Constraints expressed by experimental observables}

We rewrite the bounds on solar mixing angle in terms of 
measurable quantities. Toward the goal we note that three 
neutrino masses $m_i$ (i=1,2,3) can be expressed by the two 
$\Delta m^2$ and a remaining over-all scale $m_{H}$. 
We assign $m_{H}$ to the mass of the highest-mass state, 
$m_3$ in the normal mass hierarchy ($\Delta m^2_{23} > 0$), and 
$m_2$ in the inverted mass hierarchy ($\Delta m^2_{23} < 0$), 
respectively. 

We have argued in our previous paper \cite {MS01} that 
in a reasonable approximation one can regard $m_{H}$ 
as the observable $\langle m \rangle_{\beta}$ 
in direct mass measurements in single beta decay experiments.\footnote
{While we used the linear formula derived by Farzan, Peres and Smirnov 
\cite {FPS01}
\begin{equation}
\langle m \rangle_{\beta} = 
\frac{
\sum^{n}_{j=1}
m_j |U_{ej}|^2}
{\sum^{n}_{j=1} |U_{ej}|^2}
\label{FPS}
\end{equation}
with $n$ being the dimension of the subspace of (approximately) 
degenerate mass neutrinos, 
this point remains valid even if we use an alternative 
quadratic expression \cite{quadformula}.}
We have noticed that the identification is exact in 
two extreme cases of degenerate and hierarchical mass spectra. 
Then, the three mass eigenvalues of neutrinos can be 
represented solely by observables; 
$\Delta m^2_{atm}$, $\Delta m^2_{\odot}$, and 
$\langle m \rangle_{\beta}$ in a good approximation.

In each neutrino mass pattern, we have the expressions of 
three mass eigenvalues:

\noindent
Normal mass hierarchy ($\Delta m^2_{23} > 0$);

\begin{equation}
m_1 = \sqrt{m_H^2 - \Delta m^2_{atm} - \Delta m^2_{\odot}},
\hskip 1cm
m_2 = \sqrt{m_H^2 - \Delta m^2_{atm}},
\hskip 1cm
m_3 = m_H \simeq \langle m \rangle_{\beta}.
\end{equation}

\noindent
Inverted mass hierarchy ($\Delta m^2_{23} < 0$);

\begin{equation}
m_1 = \sqrt{m_H^2 - \Delta m^2_{\odot}},
\hskip 1cm
m_2 = m_H \simeq \langle m \rangle_{\beta}, 
\hskip 1cm
m_3 = \sqrt{m_H^2 - \Delta m^2_{atm}}.
\end{equation}

It is instructive to work out the form of constraint in the 
degenerate mass approximation, 
$m_i^2 \simeq m^2 \gg \Delta m^2_{atm}, \Delta m^2_{\odot}$.
It is easy to show that in the degenerate mass limit the bound 
(\ref{bound1}) becomes 
\begin{equation}
|\cos{2 \theta_{12}}| \leq 
\sec^2{\theta_{13}(\mbox{CH})}
\left[
\frac 
{\langle m \rangle_{\beta \beta}^{max}}{\langle m \rangle_{\beta}} + 
s^2_{13}(\mbox{CH})
\right].
\label{bound3}
\end{equation}
On the other hand, the bound (\ref{beta6}) gives the inequality 
$\langle m \rangle_{\beta} \geq 
\langle m \rangle_{\beta \beta}^{min}$ 
in the degenerate mass limit. 
(To show this one may go back to (\ref{beta6}), rather than 
using (\ref{bound2}).)

\section{Analysis of the Double Beta-Reactor Joint Constraints}

We analyze in this section the joint constraints derived 
in the foregoing sections and try to extract the implications. 
Let us start by examining the case of recent obsevation announced in 
\cite{evidence} which gives rise to 
0.11 eV $\leq \langle m \rangle_{\beta \beta} \leq$ 0.56 eV 
and 
0.05 eV $\leq \langle m \rangle_{\beta \beta} \leq$ 0.84 eV 
if $\pm 50$~\% uncertainty of the nuclear matrix elements are 
considered, each at 95~\% CL. 
In Fig.~1 we present on 
$\langle m \rangle_{\beta}$ - $\cos{2\theta_{12}}$ plane 
the constraint 
(\ref{bound1}) by the thick solid lines (solid line) and 
(\ref{bound2}) by the thick dashed line (dashed line) for cases 
with (without) uncertainty of the nuclear matrix elements, 
respectively. 
The regions surrounded by these lines are allowed.
The slope of $\langle m \rangle_{\beta}$-dependence of 
(\ref{bound2}) is so large that the dashed line looks 
like a vertical line, which implies the inequality 
$\langle m \rangle_{\beta} \geq 
\langle m \rangle_{\beta \beta}^{min}$. We have derived it 
earlier in the degenerate mass limit, but it is 
generically true if $\Delta m^2_{\odot}$ is smaller than 
other relevant mass squared scales. 
Only the case of normal mass hierarchy 
($\Delta m^2_{23} > 0$) is shown in Fig.~1; 
the case of inverted hierarchy ($\Delta m^2_{23} < 0$) gives an 
almost identical result except for a slight shift of the dashed 
line toward smaller $\langle m \rangle_{\beta}$ by 
$\simeq$ 10~\%.

Superimposed in Fig.~1 are the 95~\% CL allowed regions of 
$\cos{2 \theta_{12}}$ for the large mixing angle 
(LMA) MSW solution  
(indicated by the shaded region between thin solid lines) 
and the low (LOW) MSW solution 
(indicated by the shaded region between thin dashed lines) 
of the solar neutrino problem \cite {MSW}. 
There are several up to date global analyses of the 
solar neutrino data with similar results of allowed region of 
mixing parameters \cite {solaranalysis}. 
Therefore, we just quote the result obtaind by Krastev and Smirnov 
in the last reference in \cite {solaranalysis}.

Figure 1 illustrates that for a given value of $\cos{2 \theta_{12}}$
the single beta decay observable $\langle m \rangle_{\beta}$
has to fall into a region bounded by 
$\langle m \rangle_{\beta}^{min} \simeq 
\langle m \rangle_{\beta \beta}^{min}$ and 
$\langle m \rangle_{\beta}^{max}$, which are dictated by 
(\ref{bound2}) and (\ref{bound1}), respectively. 
Thus, we have a clear prediction for direct mass measurements using 
a single beta decay with observation of double beta decay events.
With use of the numbers given in \cite{evidence}, for example, 
the observable $\langle m \rangle_{\beta}$ must fall into 
the region 0.05 eV $\leq \langle m \rangle_{\beta} \leq$ 2 eV 
(0.11 eV $\leq \langle m \rangle_{\beta} \leq$ 1.3 eV) 
with (without) uncertainty of nuclear matrix elements 
at the best fit parameters of the LMA MSW solution. 
(The best fit value is 
$\tan^2{\theta_{12}}= 0.35$, or $\cos{2 \theta_{12}}= 0.48$
in the last reference in \cite {solaranalysis}.)
Within the allowed region the cancellation between three mass 
eigenstates can take place for appropriate values of Majorana 
phases that allow (typically) a factor of 2-3 larger values of 
$\langle m \rangle_{\beta}$ compared with the measured value of 
$\langle m \rangle_{\beta \beta}$. At around maximal mixing 
($\cos{2 \theta_{12}} \simeq 0$), which is allowed by 95~\% CL 
in the LOW solution, 
the cancellation is so efficient that much larger values of 
$\langle m \rangle_{\beta}$ is allowed.
Therefore, there are still ample room for hot dark matter mass 
neutrinos both in the LMA and the LOW solutions. 

It should be emphasized that a finite value of 
$\langle m \rangle_{\beta \beta}$ does imply a lower bound on 
$\langle m \rangle_{\beta}$, as indicated in Fig.~1;  
a vanishingly small $\langle m \rangle_{\beta}$ 
cannot be consistent with finite $\langle m \rangle_{\beta \beta}$ 
in double beta decay experiments. 
The sensitivity of the proposed KATRIN experiment is 
expected to extend to 
$\langle m \rangle_{\beta} \leq 0.3$ eV \cite {KATRIN}. 
On the other hand, the present 68~\% CL limit quoted in 
\cite {evidence} without nuclear element uncertainty is 
0.28 eV $\leq \langle m \rangle_{\beta \beta} \leq$ 0.49 eV. 
Therefore, if the limit is further tightened by additional 
data taking in the future, both experiments can become 
inconsistent, giving an another opportunity of cross checking.

In Fig.~2, we demonstrate the approximate scaling relation 
obeyed by the constraint (\ref{bound1}) by taking 
$\langle m \rangle_{\beta}/\langle m \rangle_{\beta \beta}^{max}$ 
as the abscissa in a wide range of the 
$\langle m \rangle_{\beta \beta}^{max}$ in degerarate mass region, 
0.1 eV $\lsim \langle m \rangle_{\beta \beta}^{max} \lsim$ 1 eV.
The scaling is exact in the degenerate mass limit as shown in 
(\ref{bound3}). 
The relation is useful to estimate the allowed region of 
$\langle m \rangle_{\beta}$ for a given value of 
$\langle m \rangle_{\beta \beta}^{max}$ which is not explicitly 
examined in this paper.

The most stringent bound to date on 
$\langle m \rangle_{\beta}$ is from the 
Mainz collaboration \cite {Mainz}, 
$\langle m \rangle_{\beta} \leq 2.2$ eV (95~\% CL).
(A similar bound 
$\langle m \rangle_{\beta} \leq 2.5$ eV (95~\% CL) 
is derived by the Troitsk group \cite {Troitsk}.)
As we can see in Fig.~2 that the double beta bound with the CUORE 
sensitivity region 
$\langle m \rangle_{\beta \beta} \lsim 0.3$ eV \cite {CUORE} 
becomes stronger than the Mainz bound for the LMA MSW 
solution but not for the LOW MSW solution in their 95~\% CL regions.

In Fig.~3, we present the similar allowed regions 
for hypothetical discovery of neutrinoless double beta decay 
events which would produce the experimental bounds 
0.01 eV $\leq \langle m \rangle_{\beta \beta} \leq$ 0.03 eV.
It is to examine how the constraint changes in some other  
situation of discovery with different mass parameter ranges. 
We note that even such deep region of sensitivity 
will be explored by several experiments 
\cite {GENIUS,MOON,XMASS,EXO}.

For this case, the bounds for the normal and the inverted 
mass hierarchies start to split as shown in Fig.~3. 
In the case of inverted mass hierarchy the lower bound on 
$\langle m \rangle_{\beta}$ is replaced by the trivial bound 
$\langle m \rangle_{\beta} \geq \sqrt{\Delta m^2_{atm}}$ 
which is more restrictive. The latter 
is indicated by the dash-dotted line in Fig.~3b.
It is also evident that the constraint from double beta decay 
is so stringent that the limit on $\langle m \rangle_{\beta}$ 
is tightened to be 
$\langle m \rangle_{\beta} \lsim$ 0.2 eV 
for the LMA MSW solution.

In conclusion, we have demonstrated in this and the previous papers 
the mutual intimate relationship between observation and/or 
nonobservation in   
single beta decay and neutrinoless double beta decay experiments. 
We hope that it stimulates even richer future prospects not only 
in double beta decay experiments but also in direct mass measurements 
using single beta decay. 

Finally, some remarks are in order:

\vskip 0.1cm

\noindent 
(1) If the LMA MSW solution is the case and if the KamLAND 
experiment \cite {KamLAND} 
that just started data taking can measure 
$\cos{2 \theta_{12}}$ within 10~\% level accuracy, 
the upper limit of $\langle m \rangle_{\beta}$ can be 
accurately determined with $\sim$ 20~\% accuracy.

\vskip 0.1cm

\noindent 
(2) In this paper we have derived constraints imposed on 
neutrino mixing parameters by  
observation of neutrinoless double beta decay events and 
the CHOOZ reactor bound on $|U_{e3}|^2$ 
in the generic three flavor mixing framework of neutrinos. 
Suppose that neutrinoless double beta decay events 
are confirmed to exist and the single beta decay experiments detect 
neutrino mass outside the region of the bound derived in 
this paper. What does it mean? 
It means either that double beta decay would be mediated by 
some mechanisms different from the usual one with 
Majorana neutrinos, or the three flavor mixing framework 
used in this paper is too tight. 

\vskip 0.3cm

\noindent
Nore added:
 
After submitting the first version of our paper to the 
electronic archive, we became aware of the works which 
address relatively model-independent implication of 
the results reported in \cite{evidence}, 
or critically comment on the interpretation of the events.
References \cite{list1} and \cite{list2} are the incomplete 
lists of them.

\vskip -0.4cm

\acknowledgments 

We thank Professor Klapdor-Kleingrothaus for calling 
our attention to the reference \cite{evidence}.
This work was supported by the Grant-in-Aid for Scientific Research 
in Priority Areas No. 12047222, Japan Ministry 
of Education, Culture, Sports, Science, and Technology.


\begin{figure}[ht]
\vglue 1.0cm
\hglue -1.0cm 
\includegraphics[scale=0.5]{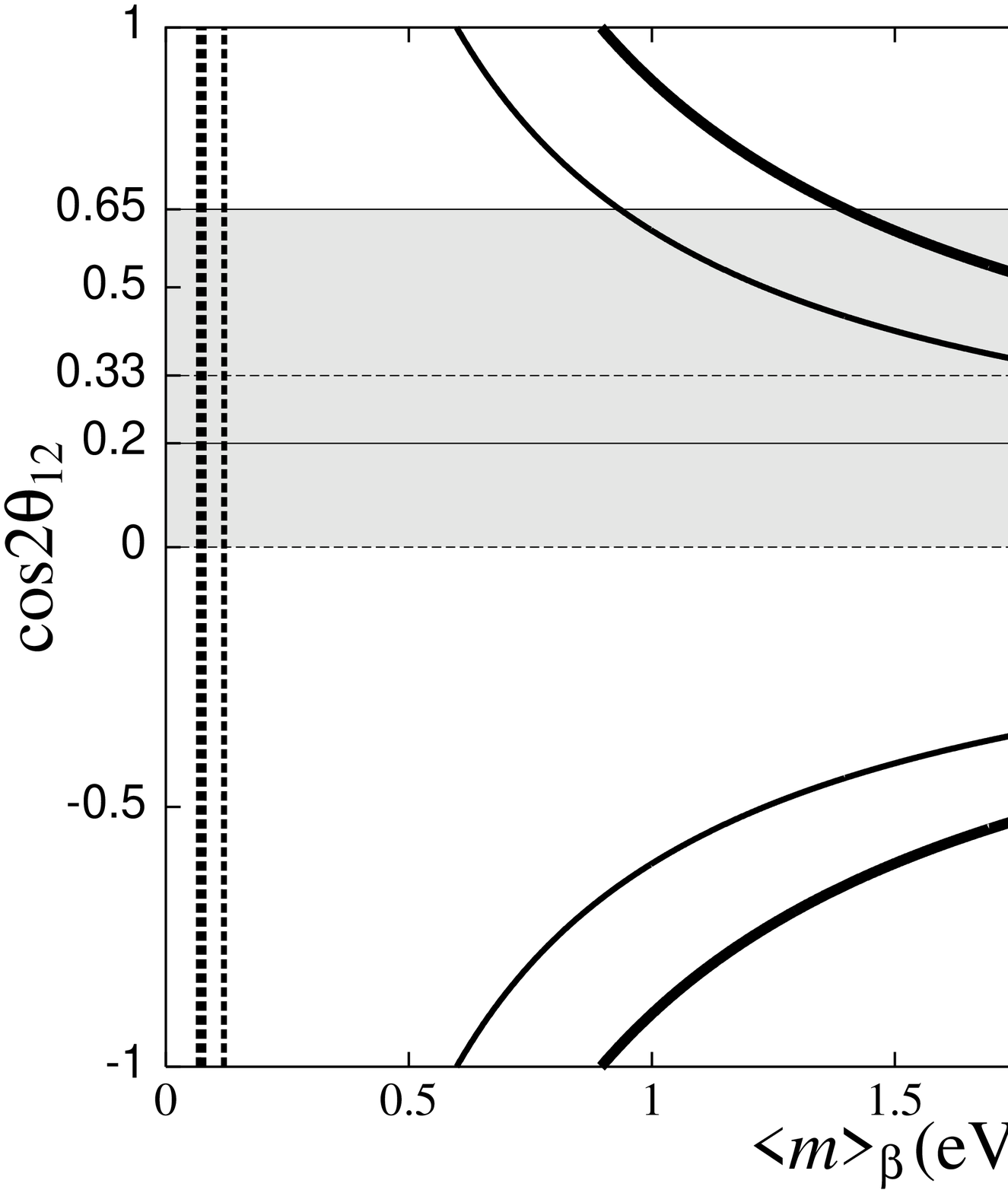}
\vglue 1.5cm 
\caption{
The constraints imposed on neutrino mixing parameters 
$\theta_{12}$ and the observable mass parameter 
$\langle m \rangle_{\beta}$ in single beta decay experiments  
by recent observation of neutrinoless double bata decay events. 
The solid and the dashed lines represent 
the bounds (\ref{bound1}) and (\ref{bound2}), respectively; 
the allowed region is inside these three lines.
The bold and the normal lines are for the ranges of mass parameter 
0.05 eV $\leq \langle m \rangle_{\beta \beta} \leq$ 0.84 eV 
and
0.11 eV $\leq \langle m \rangle_{\beta \beta} \leq$ 0.56 eV
corresponding, respectively, with and without $\pm 50$~\% 
uncertainty of nuclear matrix elements.
The mixing parameters are fixed as
$\Delta m^2_{atm} = 3 \times 10^{-3}$ eV$^2$ and 
$\Delta m^2_{\odot} = 4.8 \times 10^{-5}$ eV$^2$.
Also shown as shaded region are the allowed regions of 
$\cos{2 \theta_{12}}$ at 95~\% CL for the 
LMA (the region between thin solid lines) and 
LOW (the region between thin dashed lines) MSW solutions.}
\label{Fig1}
\end{figure}

\newpage

\begin{figure}[ht]
\vglue 2.0cm
\hglue -1.0cm 
\includegraphics[scale=0.5]{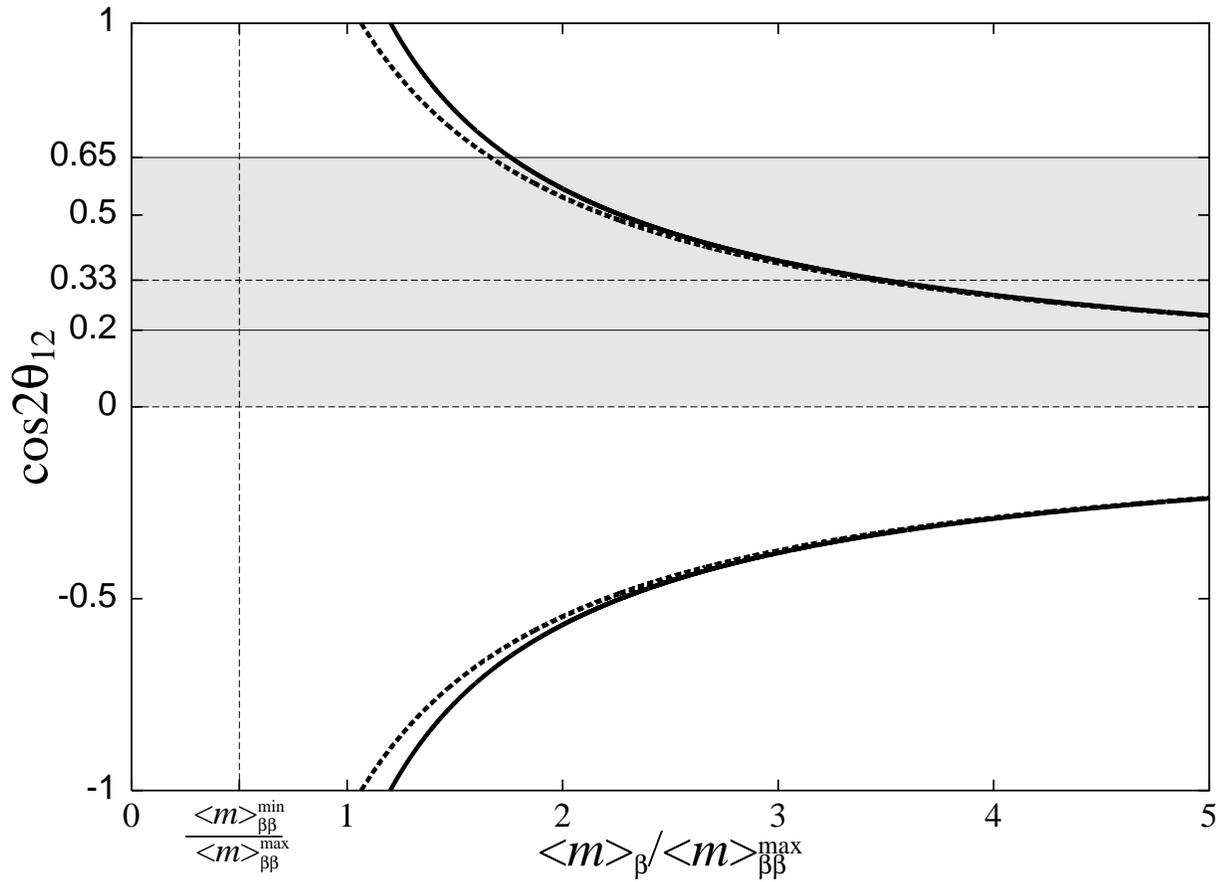}
\vglue 1.5cm 
\caption{
The approximate scaling relation obeyed by the bound (\ref{bound1}) 
for a wide range of 
$\langle m \rangle_{\beta \beta}^{max}$, 
0.1 eV (dashed lines) 
$\leq \langle m \rangle_{\beta \beta}^{max} \leq$ 
1 eV (solid lines). 
The bound (\ref{bound2}) is schematically drawn by vertical 
thin dashed line. 
The shaded regions are the same as in Fig.~1.}
\label{Fig2}
\end{figure}

\newpage

\begin{figure}[ht]
\vglue 2.0cm 
\hglue 2.5cm 
\includegraphics[scale=0.3]{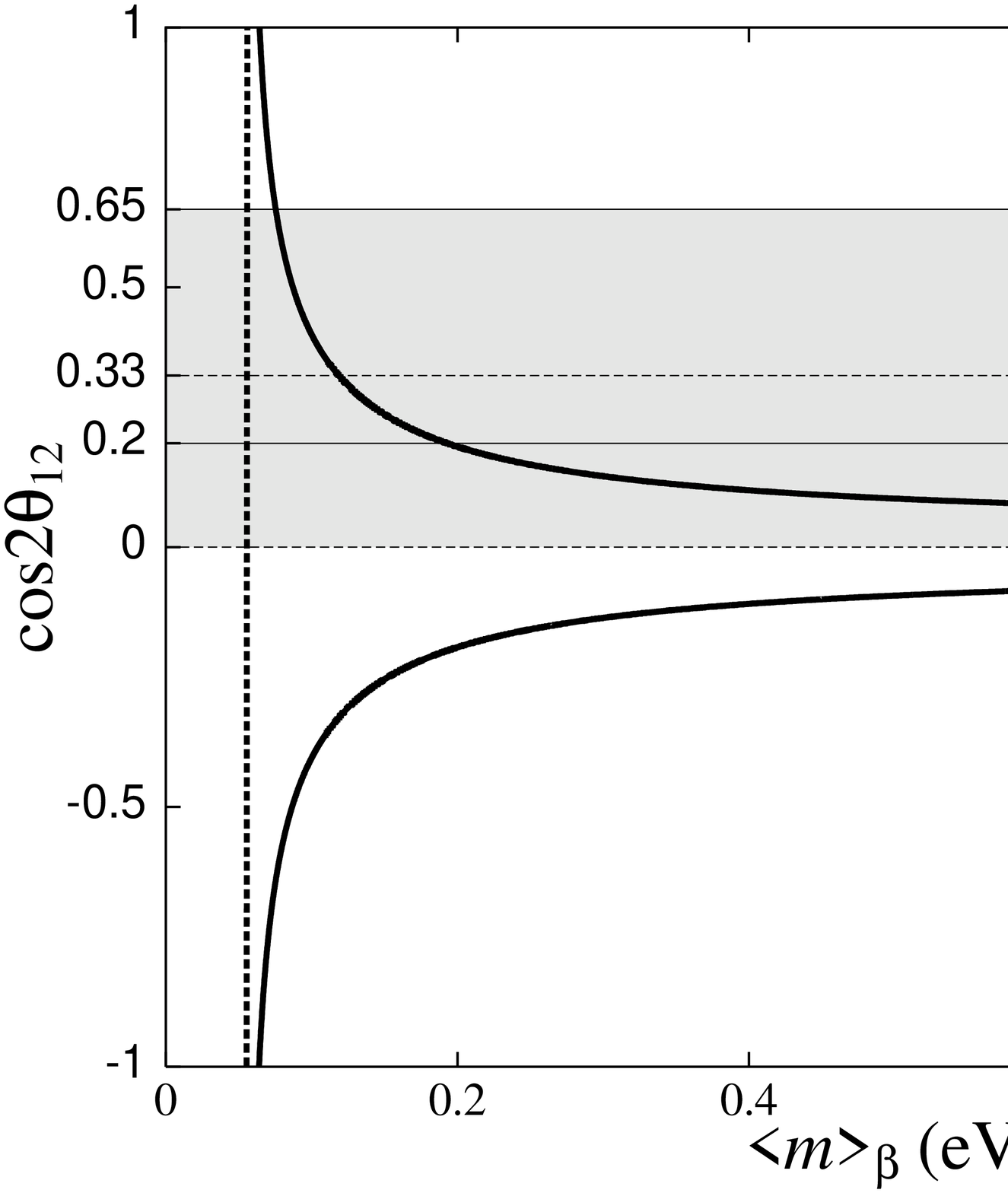}
\vglue 1.0cm
\hglue 2.5cm 
\includegraphics[scale=0.3]{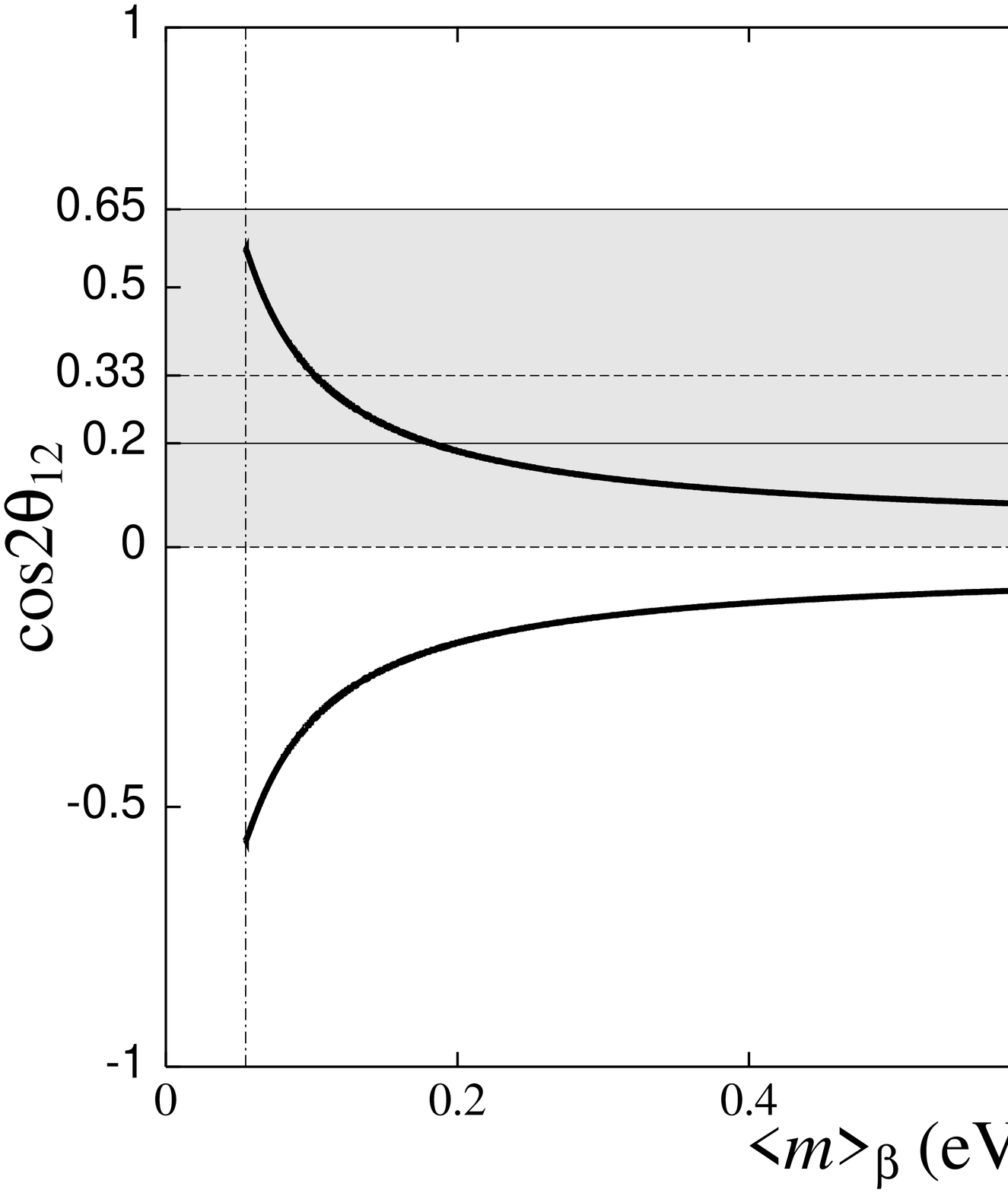}
\vglue 1.5cm 
\caption{
The same as in Fig.~2 but with assumed observed mass parameter 
$\langle m \rangle_{\beta \beta}$ in the range 
0.01 eV $\leq \langle m \rangle_{\beta \beta} \leq$ 0.03 eV.
Fig.~3a and 3b for the normal and the inverted mass hierarchies, 
respectively. 
The dash-dotted line in Fig.~3b denotes the trivial bound 
$\langle m \rangle_{\beta} \geq \sqrt{\Delta m^2_{atm}}$.}
\label{Fig3}
\end{figure}


\begin{thebibliography}{99}

\bibitem {SKatm}
Kamiokande Collaboration, Y. Fukuda {\it et al.},
Phys. Lett. {\bf B335} (1994) 237;\\
Super-Kamiokande Collaboration, Y. Fukuda {\it et al.},
Phys. Rev. Lett. {\bf 81} (1998) 1562; 
{\it ibid.} {\bf 85} (2000) 3999.


\bibitem {solar}
Homestake Collaboration, K. Lande {\it et al.},
Astrophys. J.\ {\bf 496} (1998) 505;\\ 
%
SAGE Collaboration, J.\ N.\ Abdurashitov {\it et al.},
Phys.\ Rev.\ C {\bf 60} (1999) 055801; \\
%
GALLEX Collaboration, W.\ Hampel {\it et al.}, Phys.\
Lett.\  B {\bf447} (1999) 127; \\
%
Super-Kamiokande Collaboration,  Y.\ Fukuda {\it et al.}, 
Phys. Rev. Lett. {\bf 86} (2001) 5651; 
{\it ibid.}  {\bf 86} (2001) 5656;\\
%
SNO Collaboration, Q. R. Ahmed {\it et al.}, 
Phys. Rev. Lett. {\bf 87} (2001) 071301.

\bibitem {K2K}
K2K Collaboration, S.~H.~Ahn {\it et al.},
Phys.\ Lett.\ B {\bf 511} (2001) 178;\\
See also http://neutrino.kek.jp/news/2001.07.10.News/index-e.html.

\bibitem {beta}
R.~G.~H.~Robertson and D.~A.~Knapp, 
Ann. Rev. Nucl. Part. Sci. {\bf 38} (1988) 185.


\bibitem {betabeta}
P.~Vogel, in 
{\it Current Aspects of Neutrino Physics} pp. 177, edited by 
D.O. Caldwell (Springer-Verlag, Berlin, 2001) 
[nucl-th/0005020].


\bibitem {cosmological}
D.~J.~Eisenstein, W.~Hu, and M.~Tegmark, 
Astrophys. J. {\bf 518} (1999) 2; \\
M. Fukugita, Talk at Frontiers in Particle Astrophysics and Cosmology; 
EuroConference on Neutrinos in the Universe, Lenggries, Germany, 
September 29 - October 4, 2001.


\bibitem {weiler}
T.~J.~Weiler, Phys. Rev. Lett. {\bf 49} (1982) 234;
Astroparticl Phys. {\bf 11} (1999) 303;
H.~P\"as and T.~J.~Weiler, Phys. Rev. {\bf D63} (2001) 113015;
D.~Fargion, B.~Mele, and A.~Salis, 
Astrophys. J. {\bf 517} (1999) 725;
Z.~Fador, S.~D.~Katz, and A.~Ringwald, hep-ph/0105064.


\bibitem {HeidelMos01}
Heidelberg-Moscow Collaboration, 
H. V. Klapdor-Kleingrothaus {\it et al.}, 
Eur. Phys. J. {\bf A12} (2001) 147.

\bibitem {NEMO}
C. Auger {\it et al.} (for the NEMO collaboration), 
AIP Conference Proceedings, {\bf 549} (2001) 819. 

\bibitem {GENIUS}
GENIUS Collaboration, 
H. V. Klapdor-Kleingrothaus {\it et al.}, hep-ph/9910205.


\bibitem {CUORE}
E. Fiorini {\it et al.} Phys. Rep. {\bf 307} (1998) 309; 
A. Bettini, Nucl.\ Phys.\ Proc.\ Suppl.\ {\bf 100} (2001) 332.

\bibitem {MOON}
H. Ejiri, J. Engel, R. Hazama, P. Krastev, N. Kudomi, and 
R. G. H. Robertson, Phys. Rev. Lett. {\bf 85} (2000) 2919.


\bibitem {XMASS}
S.~Moriyama, Talk at International Workshop on Technology 
and Application of Xenon Detectors (Xenon01), ICRR, Kashiwa, Japan, 
December 3-4, 2001.


\bibitem {EXO}
S.~Waldman, Talk at International Workshop on Technology
and Application of Xenon Detectors (Xenon01), ICRR, Kashiwa, Japan,
December 3-4, 2001.


\bibitem {early}
See, e.g., 
S.~T.~Petcov and A.~Yu.~Smirnov, Phys. Lett. {\bf B322} (1994) 109;\\
H.~Minakata, Phys. Rev. {\bf D52} (1995) 6630;
Phys. Lett. {\bf B356} (1995) 61; \\
S.~M.~Bilenkii {\it et al.} Phys. Rev. {\bf D54} (1996) 4432. 


\bibitem {MY97}
H. Minakata and O. Yasuda, Phys. Rev. {\bf D56} (1997) 1692; 
Nucl. Phys. {\bf B523} (1998) 597; 
O. Yasuda, in {\it Proceedings of 2nd International Conference on 
Physics Beyond the Standard Model}, edited by 
H.~V.~Klapdor-Kleingrothaus and I. Krivosheina, 
pp 223-235 (IOP Bristol 2000).

\bibitem {dblbeta1}
T. Fukuyama, K. Matsuda, and H. Nishiura, 
Mod. Phys. Lett. {\bf A13} (1998) 2279;
Phys. Rev. {\bf D57} (1998) 5844; 
{\it ibid.} {\bf D62} (2000) 093001; {\bf D63} (2000) 077301; 
{\bf D64} (2001) 013001; \\
F. Vissani, JHEP {\bf 9906} (1999) 022; \\
V. Barger and K. Whisnant Phys. Lett. {\bf B456} (1999) 194; \\
J. Ellis and S. Lola, Phys. Lett. {\bf B458} (1999) 310; \\
G. C. Branco, M. N. Rebelo, and J. I. Silva-Marcos, 
Phys. Rev. Lett. {\bf 82} (1999) 683; \\
S. M. Bilenky, C. Giunti, W. Grimus, B. Kayser, and S. T. Petcov, 
Phys. Lett. {\bf B465} (1999) 193; \\
M. Czakon, J. Gluza, and M. Zralek, Phys. Lett. {\bf B465} (1999) 211; \\
H. Georgi and S. L. Glashow, Phys. Rev. {\bf D61} (2000) 097301; \\
R. Adhikari and G. Rajasekaran, Phys. Rev. {\bf D61} (2000) 031301; \\
D. Falcone and F. Tramontano, Phys. Rev. {\bf D64} (2001) 077302. 


\bibitem {dblbeta2}
H.~V.~Klapdor-Kleingrothaus, H.~P\"as and A.~Yu.~Smirnov, 
Phys. Rev. {\bf D63} (2001) 073005; \\
S. M. Bilenky, S. Pascoli, and S. T. Petcov, 
Phys. Rev. {\bf D64} (2001) 053010; \\
S. Pascoli, S. T. Petcov, and L. Wolfenstein, 
Phys. Lett. {\bf B524} (2002) 319. 


\bibitem {MS01}
H.~Minakata and H.~Sugiyama, Phys. Lett. {\bf B526} (2002) 335, 
[hep-ph/0111269].

\bibitem {evidence}
H. V. Klapdor-Kleingrothaus, A.~Dietz, H.~L.~Harnay, and 
I.~Krivosheina, Mod. Phys. Lett. {\bf A16} (2001) 2409, 
[hep-ph/0201231].


\bibitem {CHOOZ}
CHOOZ Collaboration, M.~Apollonio {\it et al.},
Phys.\ Lett.\ B {\bf 420} (1998) 397;
{\it ibid.} B {\bf 466} (1999) 415;\\
The Palo Verde Collaboration,
F.~Boehm {\it et al.},
Phys.\ Rev.\ D {\bf 64} (2001) 112001.


\bibitem {MNS}
Z.~Maki, M.~Nakagawa and S.~Sakata,
Prog.\ Theor.\ Phys.\  {\bf 28} (1962) 870.


\bibitem {SV}
J. Schechter and J. W. F. Valle, Phys. Rev. {\bf D22} (1980) 2227;\\
S. M. Bilenky, J. Hosek, and S. T. Petcov, Phys. Lett. {\bf B94} 
(1980) 495;\\
M. Doi, T. Kotani, H. Nishiura, K. Okuda, and E. Takasugi, 
Phys. Lett. {\bf B102} (1981) 323.


\bibitem {FLMP}
G.~L.~Fogli, E.~Lisi, D.~Montanino and A.~Palazzo,
Phys.\ Rev.\ D {\bf 62} (2000) 013002.


\bibitem {FPS01}
Y. Farzan, O. L. G. Peres, and A. Yu. Smirnov, 
Nucl. Phys. {\bf B612} (2001) 59.

\bibitem {quadformula}
C.~Weinheimer {\it et al.}, Phys. Lett. {\bf B460} (1999) 219; \\ 
F.~Vissani, Nucl.\ Phys.\ Proc.\ Suppl.\  {\bf 100} (2001) 273.


\bibitem {MSW}
S.~P.~Mikheev and A.~Y.~Smirnov,
Nuovo Cim.\ C {\bf 9} (1986) 17;\\
%
L.~Wolfenstein,
Phys.\ Rev.\ D {\bf 17} (1978) 2369.

\bibitem {solaranalysis}
G. L. Fogli, E. Lisi, and D. Montanino, and A Palazo, 
Phys. Rev. {\bf D64} (2001) 093007; \\
A. Bandyopadhyay, S. Choubey, S. Goswami, and K. Kar, 
Phys. Lett. {\bf B519} (2001) 83; \\
J. N. Bahcall, M. C. Gonzalez-Garcia, and C. Pe\~na-Garay, 
JHEP {\bf 0108} (2001) 014; hep-ph/0111150; \\
P.~Cleminelli, G.~Signorelli, and A.~Strumia, hep-ph/0102234; \\
P. I. Krastev and A. Yu. Smirnov, hep-ph/0108177.


\bibitem {KATRIN}
KATRIN Collaboration, A. Osipowicz {\it et al.}, hep-ex/0109033.


\bibitem {Mainz}
J. Bonn {\it et al.}, 
Nucl.\ Phys.\ Proc.\ Suppl.\  {\bf 91} (2001) 273.

\bibitem {Troitsk}
V. M. Lobashev, {\it et al.}, 
Nucl.\ Phys.\ Proc.\ Suppl.\  {\bf 91} (2001) 280.


\bibitem {KamLAND}
Official wabpage:
http://www.awa.tohoku.ac.jp/html/KamLAND/index.html


\bibitem {list1}
H.~V.~Klapdor-Kleingrothaus and U.~Sarkar
Mod. Phys. Lett.  {\bf A16} (2001) 2469; hep-ph/0202006; \\
V.~Barger, S.~L.~Glashow, D.~Marfatia, and K.~Whisnant, hep-ph/0201262; \\
Z.-z. Xing, hep-ph/0202034; \\
N.~Haba and T.~Suzuki, hep-ph/0202143. 



\bibitem {list2}
C.~E.~Aalseth {\it et al.}, hep-ex/0202018; \\
F.~Feruglio, A.~Strumia, and F.~Vissani, hep-ph/0201291. 

\end{thebibliography}
\end{document}